\documentclass[twoside,a4paper,11pt]{sea9}
\usepackage{graphicx}
\usepackage{hyperref}
\usepackage{movie15}
\usepackage{natbib}
\begin{document}
\pagenumbering{arabic}
\pagestyle{myheadings}
\thispagestyle{empty}
{\flushleft\includegraphics[width=\textwidth,bb=58 650 590 680]{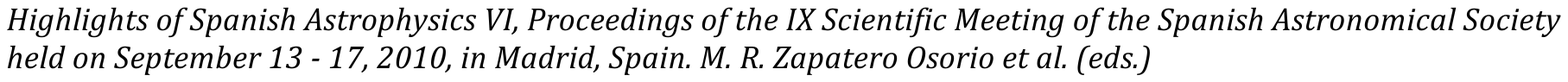}}
\vspace*{0.2cm}
\begin{flushleft}
{\bf {\LARGE
%
The Galactic O-Star Spectroscopic Survey (GOSSS)
%
}\\
\vspace*{1cm}
%
J. Ma\'{\i}z Apell\'aniz$^1$,
A. Sota$^1$,
N. R. Walborn$^2$,
E. J. Alfaro$^1$,
R. H. Barb{\'a}$^3$,
N. I. Morrell$^4$,
R. C. Gamen$^5$,
and
J. I. Arias$^3$,
%
}\\
\vspace*{0.5cm}
%
$^{1}$
Instituto de Astrof\'{\i}sica de Andaluc\'{\i}a-CSIC, Granada, Spain \\
$^{2}$
Space Telescope Science Institute, Baltimore, MD, USA \\
$^{3}$
Departamento de F\'{\i}sica, Universidad de La Serena, La Serena, Chile \\
$^{4}$
Las Campanas Observatory, La Serena, Chile \\
$^{5}$
Instituto de Astrof\'{\i}sica de La Plata (CONICET, UNLP), La Plata, Argentina
%
\end{flushleft}
%
\markboth{
The Galactic O-Star Spectroscopic Survey
}{ 
%
Ma\'{\i}z Apell\'aniz et al.
%
}
\thispagestyle{empty}
\vspace*{0.4cm}
\begin{minipage}[l]{0.09\textwidth}
\ 
\end{minipage}
\begin{minipage}[r]{0.9\textwidth}
\vspace{1cm}
\section*{Abstract}{\small
%
We present a massive spectroscopic survey of Galactic O stars, GOSSS, based on new, high signal-to-noise ratio, 
$R \sim 2500$ blue-violet digital observations from both hemispheres. The sample size and 
selection criteria; the relationship between GOSSS, the Galactic O-Star Catalog (GOSC), and three sister surveys
(OWN, IACOB, and Lucky Imaging); the current status; and our plans for the future are discussed. We also show some of our first results,
which include the new Ofc category, two new examples of Of?p stars, a new atlas for O stars, 
and the introduction of the O9.7 type for luminosity classes III to V.
Finally, our scientific objectives are discussed.
%
\normalsize}
\end{minipage}
%
%
%
\section{Introduction \label{intro}}

$\,\!$\indent O-type stars play a crucial role in the dynamic and chemical evolution of galaxies. They are the major source of ionizing 
and UV radiation and, through their huge mass-loss rates, they have a strong mechanical impact on their surroundings. Massive stars are also 
important because the nuclear reactions in their interiors create a large fraction of the heavier chemical elements. These nuclear products 
are blasted out into space in the final supernova explosions that put an end to the massive stars' lives. Despite their importance, our 
knowledge of these objects and of their evolution is still incomplete because of their relatively small numbers, average large extinctions 
caused by their concentration on the Galactic plane, and the many hidden or poorly studied multiple systems \citep{Masoetal98}.

	In 2004 we compiled the most complete Galactic O-star Catalog (GOSC) with accurate spectral types ever assembled \citep{Maizetal04b}.
GOSC was subsequently expanded from $\sim$400~objects to the current $\sim$2500 ones using stars classified as O in the literature 
\citep{Sotaetal08}. Given the variety of origins and qualities for those spectral types in the latest 
version of GOSC, they are of little use for large-scale analyses of statistical properties such as the massive-star IMF or the intensity of 
the ionizing radiation field in the Milky Way. Indeed, we have recently verified that a significant fraction of those putative O stars are 
instead B stars. Since spectroscopy is essential to study the basic properties of O stars (such as effective temperature, gravity, or mass), 
our knowledge of the massive-star population of even the solar neighborhood (within 3 kpc of the Sun) is severely limited.

\section{GOSSS description \label{description}}
	
$\,\!$\indent To remedy the incompleteness of our knowledge of the basic properties of massive stars, in 2007 we started the 
Galactic O-Star Spectroscopic Survey (GOSSS). Its immediate objective is to create the largest possible blue-violet (3900-5100 \AA) 
spectroscopic database for visually observable ($B<14$) Galactic O stars using high-S/N ($\ge$~300), intermediate spectral resolution 
($R$~$\sim$~2500) data of uniform quality starting with the 2500 objects mentioned above.

	We have currently observed $\sim$700 of those 2500 objects using three different telescopes: the 1.5~m at Observatorio de Sierra 
Nevada (OSN), the 2.5~m duPont telescope at Las Campanas Observatory, and the 3.5~m telescope at Calar Alto (CAHA). The data are processed using
a devoted pipeline (Sota \& Ma{\'\i}z Apell\'aniz, these proceedings). Our goal is to observe
at least one epoch for $\sim$2000 targets by the end of 2012 using the observatories above and possibly others. Some objects have been observed
twice or more times to compare the quality of the data obtained from different observatories (which has proven to be remarkably similar for
OSN and LCO and reasonably equivalent when comparing those two with CAHA) and to obtain different phases of the orbit for spectroscopic binaries.
We plan to keep adding 
objects to the sample based on further literature spectral types and a thorough search using data from future photometric surveys processed
through the Bayesian photometric code CHORIZOS \citep{Maiz04c}. With the exception of O9.7 stars (discussed below) we believe that the currently
observed sample is at least 99\% complete for $B<8$. Our goal is to observe all Galactic O stars down to $B=10$ and possibly deeper. 

	GOSSS is an ambitious project in terms of sample size and data quality but not in wavelength coverage, spectral resolution, or number of 
observed epochs, where other surveys provide better results. For example, its spectroscopic-binary detection capabilities and usefulness for 
detailed atmospheric modeling are limited. For those reasons, several of the authors here are also involved in two other surveys, OWN 
\citep{Barbetal10} and IACOB \citep{SimDetal10} that are observing a subset of the GOSSS stars at higher spectral resolution and multiple epochs.
OWN covers southern stars and IACOB northern stars. Also, another survey whose PI is the same one as that of GOSSS is using Lucky Imaging to 
obtain high-resolution images of another subset of GOSSS stars \citep{Maiz10}. Taken all together, the four projects will provide an unprecedented 
window into the Galactic O stars in the solar neighborhood.

	In the near future we plan to integrate the results from GOSSS into GOSC. GOSC has been recently upgraded to v2.3 and received a new URL
(\href{http:\\gosc.iaa.es}{http://gosc.iaa.es}) and interface. The information is now managed with MySQL and it is now possible to do
coordinate-based queries and to sort the results by almost any of the output columns. The output style has also changed and it now includes an
Aladin option that can be used to see the results plotted on an image. Next year (2011), GOSC will be upgraded into v3.0 and will include the new
spectral types derived from the GOSSS spectrograms. Also, the GOSSS data (and possibly those of other surveys, as well) will be made available
through the web site.

\section{First scientific results \label{results}}

$\,\!$\indent The first GOSSS scientific results were published in \citet{Walbetal10a} and Sota et al. (2010, ApJS submitted). Here are some
of our findings:

\begin{itemize}
  \item We discovered C\,{\sc iii}~$\lambda\lambda$4647-50-51 emission lines (in blend form) in otherwise normal stars of type close to O5.
        This prompted us to introduce a new category, the Ofc stars, in a manner analogous to the Of stars. Ofc stars seem to be rather
        abundant ($\sim$5\% of all O stars) and their discovery is a demonstration of the power of large uniform datasets with good S/N.
  \item The peculiar category of Of?p stars was introduced by \citet{Walb72} and for almost 40 years it included just three Galactic members:
        HD~108, HD~148\,937, and HD~191\,612. The interest in the class has been recently revived by the discovery of magnetic fields in all of
	them \citep{Donaetal06,Martetal10,Wadeetal10}. In the GOSSS data we have found two new Of?p stars, NGC 1624-2 and 
	CPD -28$^{\rm o}$ 2561, raising their total Galactic number to five.
  \item We have presented a high-quality O-star atlas that can be used for spectral classification. The atlas has higher S/N and fewer gaps than
        previous examples and is accompanied by MGB, a software that can be used to compare observed data with the standards, thus facilitating 
        classification.
  \item We have adopted a uniform horizontal classification criterion for late O stars based on the 
        He\,{\sc ii}~$\lambda$4542/He\,{\sc i}~$\lambda$4388 and He\,{\sc ii}~$\lambda$4200/He\,{\sc i}~$\lambda$4144 ratios. As a result, the O9.7 type
	has been extended to luminosity classes V to III (previously, it was only defined for classes II to Ia). This change in definition
        implies that many former B0 V to III stars are now of type O9.7, thus increasing the total number of O stars by
	5-10\%. One object that is now considered to be O9.7 V is $\upsilon$ Ori, the former B0 V standard. 
  \item With respect to the vertical classification criteria, the new data have allowed us to distinguish luminosity class IV for spectral
        classes O6--O8 for the first time and to discover several new examples of type O~Vz.
  \item Other changes to the classification criteria include the elimination of the ((n)) qualifier (line width, existing cases have become either (n) or normal) 
        and of the f+ qualifier (presence of Si\,{\sc iv}~$\lambda$4089-4116 emission).
  \item We have discovered some interesting correlations between extinction, DIB intensity and the equivalent width of the interstellar
        Ca\,{\sc ii}~$\lambda$3934 absorption line (Penad\'es Ordaz et al., these proceedings).
\end{itemize}

\section{Objectives \label{objectives}}

$\,\!$\indent 	The final objectives of GOSSS run along five different research lines:

	{\it 1. The fundamental spectral morphology of Galactic O stars.} The previously published libraries of digital spectrograms of O stars at 
R~$\sim$~2500, though detailed and complete in terms of spectral type coverage, are far from exhaustive since they include $<$100 O stars in total. 
Furthermore, those libraries include only a few northern stars and the quality of their data can be significantly improved with modern CCD detectors. 
An improvement in size (from 100 to 2000 O stars), quality (from S/N of 50 to 300), and uniformity of a spectral library similar to the one in this
project has always led in the past to new discoveries in the observed spectral morphologies. Those discoveries have produced, first, changes in the 
classification criteria and, later on, significant advances in the understanding of the underlying physical principles through comparison with atmosphere 
models. We expect those two steps to take place in this project as well and some examples have been presented in the previous section.

	{\it 2. The multiplicity of massive stars.} Most, if not all, massive stars are born in multiple systems \citep{Masoetal98}. This represents both a
blessing and a curse for their study. It is a blessing because, at least in principle, it allows the measurement of their masses. The curse comes from the 
difficulty in doing so: long-period (thousands of years or more) Galactic massive binaries usually require high-resolution imaging or interferometry while 
short-period (days to months) ones require time-consuming multiple-epoch spectroscopy. In-between objects, those with unfavorable orientations, and 
high-order multiples with complex hierarchical orbits remain undetected (hence, with incorrect estimated masses) or with only uncertain values, thus 
hampering the derivation of the IMF. GOSSS, in conjunction with the OWN, IACOB, and our high-resolution imaging survey will attack this problem in a threefold way: 
[1] observing with multiple-epoch spectroscopy at high and intermediate resolution (with this approach we are following the orbits of several tens of massive
multiple systems, many of them previously unknown); [2] observing with high-resolution imaging; and [3] compiling information from the literature. This 
should yield the most complete ever study of the multiplicity of Galactic massive stars. Our preliminary OWN results are spectacular: out of 240 O and WN stars 
we have detected radial velocity variations above 10 km/s for more than 100 of them and for 26 objects in the sample we already have calculated spectroscopic 
binary orbits \citep{Barbetal10}. This raises the fraction of spectroscopic multiples among massive stars from $\sim$20\% to $\sim$50\%. Furthermore, the period 
distribution appears to be very different for O and B stars, 
something that will require further studies to be understood. An example of the importance of high-resolution imaging studies is shown in \citet{Maizetal07}.
 
	{\it 3. The optical-IR extinction law and the nature of the DIBs.} Nowadays, the most used interstellar extinction laws are those from 
\citet{Cardetal89}. The situation with their quality is similar to that of O-star spectral libraries mentioned in the first research line: they were derived 
with old data and a small sample (less than 30 stars, all with $E(B-V)<1.3$) that can be greatly improved with modern means, thus overcoming some of the 
known issues with the Cardelli laws. Here, we propose doing just that, combining the optical-NIR photometry for the stars in GOSC with our newly derived 
spectral types using CHORIZOS \citep{Maiz04c} and the newly recalibrated photometric zero points \citep{Maiz05b,Maiz06a,Maiz07a}. The newly derived 
extinction laws will have variable NIR slopes. The combination of a large library of high S/N optical O-star spectra (from GOSSS, OWN, and IACOB) 
with an improved knowledge of the amount and type of extinction for each star will also allows us to analyze the decades-old problem of the nature of the 
DIBs \citep{Herb95} with a fresh impulse.

	{\it 4. The spatial distribution of massive stars and dust within 3 kpc of the Sun.} Most recent studies of the spatial distribution of Galactic 
extinction have concentrated on either [a] the diffuse component at high latitudes and low extinction (e.g. \citealt{Jurietal08}) or [b] the high extinction 
regions located behind molecular clouds or at large distances (e.g. \citealt{Alveetal01}) because those studies can be effectively carried out using large 
photometric surveys where the target stars are of A and later type. On the other hand, the once popular surveys based on OB-star spectroscopy 
(e.g. \citealt{Necketal80}) have faded out of style, leaving the region within 3 kpc of the Sun mostly unstudied even though the quality of the available 
data is much better than 30 years ago. Our data, with the aid of completeness information from 2MASS and (in the future) WISE, should cover this important 
gap and provide the best available information on the spatial distribution of massive stars and dust within 3 kpc of the Sun. 

	{\it 5. The massive end of the IMF in the solar neighborhood.} Once the four previous research lines have reached their major objectives (accurate 
spectral types, characterization of multiple systems, and precise extinction corrections and spectroscopic distances for our sample) we will attack one of 
the current holy grails of Galactic research, the IMF for massive stars, with data of unprecedented quality and overcoming many of the biases present in 
previous attempts. Our goal here is to redo the work of \citet{Massetal95a} with better data, a larger sample, and improved numerical techniques.

\small  
%
\section*{Acknowledgments}   
%
Support for this work was provided by [a] the Spanish Ministerio de Ciencia e Innovaci\'on through 
grants AYA2007-64052 and AYA2010-17631, the Ram\'on y Cajal Fellowship program, and FEDER funds; [b] the Junta de Andaluc\'{\i}a
grant P08-TIC-4075; and [c] NASA through grants GO-10205, GO-10602, and GO-10898 from STScI, which is operated by AURA Inc., under 
NASA contract NAS~5-26555. This research has made extensive use of Aladin \citep{Bonnetal00};
the SIMBAD database, operated at CDS, Strasbourg, France; and the WDS Catalog 
maintained at USNO \citep{Masoetal01}.

%
%
%
%
\bibliographystyle{aj}
\small
\bibliography{general}

\end{document}